\documentclass[11pt]{revtex4-2}

\usepackage[dvips]{color}
\usepackage{epsfig}
\usepackage{amsmath}
\usepackage{graphicx}
\usepackage{subfigure}
\usepackage[dvips]{color}
\usepackage{epsfig}
\usepackage{amsmath}
\usepackage{graphicx}
\usepackage{subfigure}
\begin{document}
\title{ Bouncing universe for deformed non-minimally  coupled inflation model}
\author{S. Upadhyay}
\email{sudhakerupadhyay@gmail.com}
 
\affiliation{Department of Physics, K.L.S. College, Magadh University, Nawada-805110, Bihar,  India}

\begin{abstract}
In this paper, we consider a non-minimally coupled gravity model to study the bouncing universe. The holographic principle has various effects on the bouncing universe. We choose some suitable new variables and achieve the new Hamiltonian and Lagrangian which have harmonic oscillator form. The corresponding Lagrangian is deformed by non-commutative geometry. In order to have a solution for the bouncing universe we specify the potential in the equation state. In that case, we draw the equation of state in terms of time and show that the equation of state crosses $-1$. Such bouncing behavior leads us to apply some conditions on $\theta$ and $\beta$ from non-commutative geometry. Here, we can also check the system's stability due to the deformation of the non-minimally coupled gravity model. In order to examine the
stability of the system we obtain the variation of pressure with respect to density energy. Also, we draw the variation of pressure with respect to energy density and show the stability condition.
\end{abstract}

\maketitle
\section{Introduction}
The moving from the accelerated collapse era to the expanding era without a singularity lead us to have a bouncing universe. The bounce realization due to the holographic principle in the early universe studied recently \cite{ond}. The holographic
dark energy  plays a crucial role in
describing the late-time universe, and found in agreement
with observations \cite{ond1,ond2,ond3,ond4,ond5}.
In that case, many cosmological solutions for the bouncing universe tested in the last decades \cite{1}. The subject of bouncing investigated in several model as brane-world
scenarios of cosmology. Here, if the bulk space is taken to be a charged AdS
black hole, then we have the bouncing universe \cite{2}. In this case,
brane makes a contracting to expanding phase transition. As we know from the singularity
theorem in four-dimensions, such a bounce cannot occur because of the certain energy condition. Hence, in order to produce the bouncing universe, it is necessary that the bulk geometry contribute a negative energy density to the effective stress-energy tensor of the brane configuration \cite{3}. As we know, several authors give further developments for these bouncing brane-world scenario \cite{4,5,6,7,8,9,10}. Now, it is believed that the universe is in accelerated expansion phase. For this cosmic acceleration, we have seen some observations from type Ia supernovae \cite{14,15,16} in associated with Large Scale Structure (LSS) \cite{17,18} and Cosmic Microwave Background (CMB) anisotropy \cite{19,20}. On the other hand the dark energy with negative pressure lead us to this cosmic acceleration. In order to have such phenomena, the theories  trying to modify Einstein equations. On the other hand, in modern cosmology, the dark energy can be explained by  equation of state parameter as  $\omega=\frac{p}{\rho}$, where $p$ and $\rho$
are pressure and energy density, respectively. In that case, for the  accelerated expansion the equation of state dark energy must be satisfied by $\omega< -\frac{1}{3}$.\\
As we know, there are two puzzles from cosmological constant as dark energy. One of them is that, cosmological constant is about 120 orders of magnitude smaller than its natural expectation. The other is cosmological coincidence. In order to solve these problems, various model of dark energy have been introduced such as quintessence \cite{21,22,23} or k-essence \cite{k}, scaler-field dark energy model including tachyon \cite{24,25}, ghost condensate \cite{26,27} and quintom \cite{28,29,30,31,32,34,35,36}. Also there is a unified dark energy-dark matter model based on Chaplygin gas equation of state and its extensions \cite{C1,C2,C3,C4,C5,C6,C7,C8,C9,C10}. We note here the analysis of the dark energy properties from recent observation mildly favour model with $\omega$ crossing $-1$ in the near past. In that case there are other proposals to describe accelerating expansion of the universe include interacting dark energy models \cite{37,40}, brane-world models \cite{41,42}, and holographic dark energy models \cite{43,44,45,49}. The difficulty of realizing $\omega$ crossing over $-1$ in quintessence and phantom-like models for the first time have been shown by Refs. \cite{29,32,34}. Because when the equation of state $\omega$ approaches to $-1$, the dark energy perturbation would be divergent \cite{51,52,56,57}. The most important point here is that the quintom scenario of dark energy very well understand the nature of dark energy with $\omega$ across $-1$ \cite{58,59}.\\
All above physical information give us motivation to work non-minimally coupled gravity with deformed phase space. As we know, such non-minimal coupling is very successful \cite{60,61}.  The simple example of such model is Higgs boson which act the inflation field in the form of $\xi\varphi^{2}R$. In that case $\varphi$ and $R$ are Higgs field and Ricci scaler respectively. Here, we are going to study the dark energy and bouncing universe $\omega$ with usual metric formalism in non-minimally coupling model. We deformed the non-minimally coupling to gravity model. Then, we compare the results of deformed and non-deformed corresponding model to each other. Here, we note that the bouncing universe and dark energy in non-minimally coupled gravity with deformed phase space lead us to have some new results. Here, another deformed approach will be Finsler geometry. In future one can realize the relation between the above deformed method, non-commutative geometry and Finsler geometry.\\
So, in section \ref{sec2}, we introduce the non-minimal coupling to the gravity model. Also, we take Friedmann-Roberson-Walker (FRW)  background and write the corresponding Lagrangian. In section \ref{sec3}, first of all we assume $\xi=\frac{-1}{6}$, $\kappa=0$ and write the corresponding Lagrangian. In order to deform such a model we chose some changes of variables. Also, again we obtain the Lagrangian and Hamiltonian of the non-minimal model by new variables. The corresponding Hamiltonian will be form of harmonic oscillator, and it is useful to investigate the gravitational theories. On the other hand, we review some non-commutative (NC) geometries \cite{62}. We transform the classical phase space variables of $x_{i}$, $y_{i}$, $P_{x_{i}}$ and $P_{y_{i}}$. Also, we achieve some deformed algebra for the new variables which is satisfied by anti-commutative relations. Here, the NC geometry helps us to write the Hamiltonian and Lagrangian for the non-minimal coupling to gravity model with the new variable. In section \ref{sec4}, we have some cosmology solutions for the deformed non-minimally model and investigate the dark energy and bouncing universe. The equation of state plays important role in the bouncing universe. Here also, we have some figures for the equation of state cross $-1$. In the  last section, we have some results and suggestions for the deformed model in the bouncing universe.
\section{Lagrangian of  non-minimal coupling to gravity}\label{sec2}
Now, we consider the general model of inflation with non-minimal coupling which has a following action,
\begin{eqnarray}
S=\int d^{4}x \sqrt{-g}\left[\frac{1}{2}(M^{2}+\xi \varphi^{2})g^{\mu \nu}R_{\mu \nu}-\frac{1}{2}g^{\mu \nu}\partial_{\mu}\varphi \partial_{\nu}\varphi-V(\varphi)\right],\label{1}
\end{eqnarray}
where $M$, $R_{\mu \nu}$, $\xi$ and $\varphi$ are mass scales, Ricci tensor, dimensionless coupling constant and the inflation field respectively. If we back to the Einstein frame, we need the following conformal transformation,
\begin{equation}\label{2}
g_{\mu\nu}\rightarrow\Omega(\varphi)^{-1}g_{\mu\nu},
\end{equation}
where
\begin{equation}
\Omega(\varphi)=\frac{M^{2}+\xi\varphi^{2}}{M_{p}^{2}}.
\end{equation}
We do not need such transformation because we take general form of action (\ref{1}), and employ the FRW metric background with $\kappa=0$. Here, we try first to study the bouncing universe and dark energy to usual non-minimal coupling model (\ref{1}).
In order to have some cosmology solutions, one can write the corresponding Lagrangian with the following equation,
\begin{equation}\label{4}
\mathcal{L}=3M^{2}(\kappa a-a\dot{a}^{2})+3\xi(-a\dot{a}^{2}\varphi^{2}-2a^{2}\dot{a}\varphi\dot{\varphi}+\kappa a\varphi^{2})+\frac{1}{2}a^{3}\dot{\varphi}^{2}-V(\varphi)a^{3},
\end{equation}
where we used,
\begin{eqnarray}\label{3}
\sqrt{-g}&=&a^{3},\nonumber\\
R&=&6(\frac{\ddot{a}}{a}+\frac{\kappa+\dot{a}^{2}}{a^{2}}).
\end{eqnarray}
The equation (\ref{4}) is the Lagrangian of non-minimal coupling to gravity. In the next section, we are going to take the above Lagrangian and apply some transformations to the phase space variables.
\section{Deformation of non-minimal coupling to gravity model}\label{sec3}
Now, we take the equation (\ref{4}) and assume $\xi=\frac{-1}{6}$ and $\kappa=0$, to rewrite the Lagrangian as following,
\begin{equation}\label{5}
\mathcal{L}=3M^{2}(-a\dot{a}^{2})+\frac{1}{2}(a\dot{a}^{2}\varphi^{2}+2a^{2}\dot{a}\varphi\dot{\varphi}+a^{3}\dot{\varphi}^{2})-V(\varphi)a^{3}.
\end{equation}
Here, we choose the following change of variables,
\begin{equation}\label{6}
\begin{array}{ccc}
x_{1}=a\sinh (\frac{\varphi a}{\sqrt{2}}), & x_{2}=a\cosh \sqrt{3a}M, & x_{3}=\sinh (\sqrt{V(\varphi)}a^{\frac{3}{2}}), \\
y_{1}=a\cosh (\frac{\varphi a}{\sqrt{2}}), & y_{2}=a\sinh \sqrt{3a}M, & y_{3}=\cosh (\sqrt{V(\varphi)}a^{\frac{3}{2}}),
\end{array}
\end{equation}
and
\begin{equation}\label{7}
\begin{array}{c}
x_{4}+y_{4}=1+\sqrt{V(\varphi)}a^{\frac{3}{2}},\\
x_{4}-y_{4}=1-\sqrt{V(\varphi)}a^{\frac{3}{2}}.
\end{array}
\end{equation}
We put the equations (\ref{6}) and (\ref{7}) into the equation (\ref{5}) and obtain the following Lagrangian,
\begin{equation}\label{8}
\mathcal{L}=\sum\limits_{i=1}^4 ((\dot{x}_{i}^{2}-\dot{y}_{i}^{2})+(x_{i}^{2}-y_{i}^{2})).
\end{equation}
By using the equation $\mathcal{H}=P_{x_{i}}\dot{x}_{i}+P_{y_{i}}\dot{y}_{i}-\mathcal{L}$ one can write the corresponding Hamiltonian as,
\begin{equation}\label{9}
\mathcal{H}=\frac{1}{4}\sum\limits_{i=1}^4 ((P_{x_{i}}^{2}-P_{y_{i}}^{2})+\omega_{i}^{2}(x_{i}^{2}-y_{i}^{2})),
\end{equation}
where
\begin{equation}\label{10}
\begin{array}{c}
P_{x_{i}}=\frac{\partial\mathcal{L}}{\partial\dot{x_{i}}}=2\dot{x}_{i},\\
P_{y_{i}}=\frac{\partial\mathcal{L}}{\partial\dot{y_{i}}}=-2\dot{y}_{i}.
\end{array}
\end{equation}
It is obvious that the Hamiltonian (\ref{9}) has a harmonic oscillator form with the following frequency,
\begin{eqnarray}\label{11}
\omega_{i}^{2}=-4.
\end{eqnarray}
Here, we note that the Hamiltonian (\ref{9}) is very useful to obtain the solution of any cosmological system. Now, we are going to deform the non-minimal coupling to the gravity system. In order to do such a process, we need to give some transformations to the classical phase space. So, the first, we review such transformation as an NC geometry. Therefore, before applying NC geometry to the new Hamiltonian (\ref{9}), we give some explanation of such geometry. As we know, in commutative space we have the following Poisson brackets,
\begin{equation}\label{12}
\{x_{i},x_{j}\}=0,\quad\{P_{x_{i}},P_{x_{j}}\}=0,\quad\{x_{i},P_{x_{j}}\}=\delta_{ij},
\end{equation}
where $i=1, 2$. But in NC geometry the Moyal product plays an important role in the deformation phase space variables. Also, the quantum effects can be dissolved by the following Moyal brackets,
\begin{equation}\label{13}
\{f,g\}_{\alpha}=f\star_{\alpha}g-g\star_{\alpha}f.
\end{equation}
The brackets (\ref{13}) are based on the Moyal product which is given by,
\begin{equation}\label{14}
(f\star_{\alpha}g)(x)=exp[\frac{1}{2}\alpha^{ab}\partial_{a}^{(1)}\partial_{b}^{(2)}]f(x_{1})g(x_{2})|_{x_{1}=x_{2}=x}.
\end{equation}
By using the equations (\ref{13}) and (\ref{14}), one can find the following relation,
\begin{equation}\label{15}
\{x_{i},x_{j}\}_{\alpha}=\theta_{ij},\quad\{x_{i},P_{j}\}_{\alpha}=\delta_{ij}+\sigma_{ij},\quad\{P_{i},P_{j}\}=\beta_{ij}.
\end{equation}
So, in that case the transformations of the classical phase space variables are given by,
\begin{equation}\label{16}
\hat{x}_{i}=x_{i}+\frac{\theta}{2}P_{y_{i}},\quad \hat{y}_{i}=y_{i}-\frac{\theta}{2}P_{x_{i}},\quad \hat{P}_{x_{i}}=P_{x_{i}}-\frac{\beta}{2}y_{i},
\quad \hat{P}_{y_{i}}=P_{y_{i}}+\frac{\beta}{2}x_{i},\\
\end{equation}
and
\begin{equation}\label{17}
\{\hat{y},\hat{x}\}=\theta,\quad\{\hat{x},\hat{P}_{x}\}=\{\hat{y},\hat{P}_{y}\}=1+\sigma,\quad\{\hat{P}_{y},\hat{P}_{x}\}=\beta,
\end{equation}
where $\sigma=\frac{\beta\theta}{2}$. In order to construct the
deformed Hamiltonian for the non-minimal coupling to gravity model, we apply the new variables (\ref{16}) on the Hamiltonian (\ref{9}). Hence, the Hamiltonian (\ref{9}) in deformed form will be as,
\begin{equation}\label{18}
\hat{\mathcal{H}}=\frac{1}{4}\sum_{i=1}^{4}((P_{x_{i}}^{2}-P_{y_{i}}^{2})-\gamma_{i}^{2}(y_{i}P_{x_{i}}+x_{i}P_{y_{i}})+
{\tilde{\omega}_{i}^{2}}(x_{i}^{2}-y_{i}^{2})),
\end{equation}
where $\omega_{i}^{2}$ can be changed by the following equation,
\begin{equation}\label{19}
{\tilde{\omega}_{i}}^{2}=\frac{{\omega_{i}}^{2}-\frac{\beta^{2}}{4}}{1-{\omega_{i}}^{2}
\frac{\theta^{2}}{4}},\quad
{\gamma_{i}}^{2}=\frac{\beta-{\omega_{i}}^{2}\theta}{1-{\omega_{i}}^{2}\frac{\theta^{2}}{4}}.
\end{equation}
Here, two parameters of NC geometry as $\theta$ and $\beta$ play important role for the deformation of theory. When we investigate the bouncing universe, we will explain the role of two parameters $\theta$ and $\beta$ with some figures.
The above deformed Hamiltonian lead us to have a new Lagrangian which is deformed form of original Lagrangian of non-minimal coupling to gravity model.\\
Now, for the corresponding model, we arrange the deformed Lagrangian as,
\begin{equation}\label{20}
\hat{\mathcal{L}}=\frac{1}{4}\sum_{i=1}^{4}((P_{x_{i}}^{2}-P_{y_{i}}^{2})-{\hat{\gamma}_{i}^{2}}(y_{i}P_{x_{i}}+x_{i}P_{y_{i}})-
{\hat{\omega}_{i}^{2}}(x_{i}^{2}-y_{i}^{2})),
\end{equation}
where
\begin{equation}\label{21}
{\hat{\omega}_{i}}^{2}=\frac{{\omega_{i}}^{2}+\frac{\beta^{2}}{4}}{1+{\omega_{i}}^{2}
\frac{\theta^{2}}{4}},\quad
{\hat{\gamma}_{i}}^{2}=\frac{\beta+{\omega_{i}}^{2}\theta}{1+{\omega_{i}}^{2}\frac{\theta^{2}}{4}}.
\end{equation}
We use the equations (\ref{10}) and (\ref{20}) to obtain the deformed Lagrangian with respect to $a$, $\dot{a}$, $\varphi$, $\dot{\varphi}$ and $V(\varphi)$ which is given by,
\begin{eqnarray}\label{22}
\hat{\mathcal{L}}=-3M^{2}a\dot{a}^{2}+\frac{1}{2}(a\dot{a}^{2}\varphi^{2}+2a^{2}\dot{a}\varphi\dot{\varphi}+
a^{3}\dot{\varphi}^{2})-\lambda_{1}^{2}V(\varphi)a^{3}+\lambda_{2}^{2}(\frac{\dot{\varphi}a^{3}}{\sqrt{2}}+
\frac{a^{2}\dot{a}\varphi}{\sqrt{2}}-\sqrt{3}M\dot{a}a^{\frac{3}{2}}),
\end{eqnarray}
where
\begin{equation}\label{23}
\lambda_{1}^{2}=\frac{1-\frac{\beta^{2}}{16}}{1+\theta^{2}},\quad
\lambda_{2}^{2}=\frac{2\theta-\frac{1}{2}\beta}{1-\theta^{2}}.
\end{equation}
The above Lagrangian is deformed shape of original Lagrangian given by the equation (\ref{5}). So the above deformed Lagrangian shows us how the NC parameters effect to the corresponding theory. We note here, for the non-minimal model there were several papers for investigating the bouncing universe, and they have shown that the equation of state cross $-1$. But, we use the deformed non-minimal coupling to gravity model and obtain the equation of state. The corresponding equation of state for the deformed case lead us to investigate the bouncing universe.
\section{Bouncing universe of deformed non-minimal coupling to gravity model}\label{sec4}
According to the conservation of energy,
\begin{equation}\label{24}
E_{\hat{\mathcal{L}}}=P_{a}\dot{a}+P_{\varphi}\dot{\varphi}-\hat{\mathcal{L}}
\end{equation}
where
\begin{equation}\label{25}
P_{a}=\frac{\partial\hat{\mathcal{L}}}{\partial\dot{a}},\quad
P_{\varphi}=\frac{\partial\hat{\mathcal{L}}}{\partial\dot{\varphi}}
\end{equation}
Here, by using definition of Hubble parameter $H=\frac{\dot{a}}{a}$, and assumption $E_{\hat{\mathcal{L}}}\equiv 0$, we will arrive at,
\begin{equation}\label{26}
H^{2}=\frac{1}{6M^{2}}(H\varphi+\dot{\varphi})^{2}+\frac{\lambda_{1}^{2}}{3M^{2}}V(\varphi).
\end{equation}
The Euler-Lagrangian equation help us to obtain the following equations,
\begin{equation}\label{27}
\ddot{\varphi}+3H\dot{\varphi}+(2H^{2}+\dot{H})\varphi+\sqrt{2}\lambda_{2}^{2}H+\lambda_{1}^{2}V^{'}(\varphi)=0,
\end{equation}
and
\begin{equation}\label{28}
\dot{H}=-\frac{2(H\varphi+\dot{\varphi})^{2}+\sqrt{2}\lambda_{2}^{2}(H\varphi+\dot{\varphi})+
\lambda_{1}^{2}V^{'}(\varphi)\varphi}{6M^{2}}.
\end{equation}
And then, by using $\dot{H}=-\frac{1}{2}(\rho+p)$, one can rewrite the equation (\ref{28}) as,
\begin{equation}\label{29}
p+\rho=\frac{\frac{2}{3}(H\varphi+\dot{\varphi})^{2}+\frac{\sqrt{2}}{3}\lambda_{2}^{2}(H\varphi+\dot{\varphi})
+\frac{\lambda_{1}^{2}}{3}V^{'}(\varphi)\varphi}{M^{2}}.
\end{equation}
Then, by using $p+\rho=(1+\omega)\rho$ and assuming $\dot{\varphi}=0$ and $\omega= -1$, from the equation (\ref{23}), and $\theta^2=\beta^2=0$, one can obtain the following equation,
\begin{equation}\label{30}
\frac{2}{3}(H\varphi)^{2}+\frac{\sqrt{2}}{3}(2\theta-\frac{1}{2}\beta)(H\varphi)
+\frac{1}{3}V^{'}(\varphi)\varphi=0.
\end{equation}
The bouncing behavior of deformed of non-minimally coupled to gravity lead us to arrange the suitable potential of system. For this reason, we need to check the necessary condition for the bouncing universe, such condition help us to obtain the corresponding potential. As we know, during the contracting phase, the scale factor $a(t)$ is decreasing, it means $\dot{a}<0$, For the expanding phase, we have $\dot{a}>0$. Also, we note here for a period of time at bouncing point and around this point we have $\dot{a}=0$ and $\ddot{a}>0$ respectively. At the bouncing point, we have a transition from $H<0$ to $H\geq0$ \cite{C10}.\\

\begin{figure}[h]
\hspace*{1cm}
\begin{center}
\epsfig{file=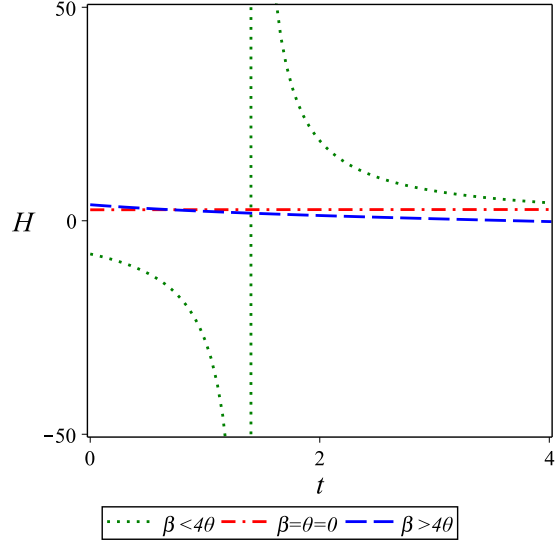,width=7.5cm}
 \caption{$H$ with respect to $t$, where $M=1$, $\varphi=0.5$, $\theta=2$.}
\end{center}
\end{figure}
In the Fig. 1 we show numerical solution of the equation (\ref{28}) where $H < 0$ to $H > 0$ transition illustrated. We can see modified parameters are necessary to have bouncing universe. Green dotted line of the Fig. 1 show that at the special time the Hubble expansion parameter change the sign from plus to minus. In absence of modification, the Hubble expansion parameter is a constant.\\
We can also find analytical expression, in such case the suitable bouncing universe can obtained by,
\begin{equation}\label{31}
H=\frac{-\frac{\sqrt{2}}{2}\lambda_{2}^{2}\pm\sqrt{\frac{1}{2}\lambda_{2}^{4}-2\lambda_{1}^{2}V^{'}
(\varphi)\varphi}}{2\varphi}.
\end{equation}
In order to have bouncing universe we have to consider three cases, which help us to obtain the exactly potential for the non-minimally coupled to gravity model.\\
The first case: if we assume $H=0$, one can obtain $V^{'}(\varphi)=0$, therefore $V(\varphi)=constant$.\\
The second case: if the Hubble parameter $H$ is positive, then $V^{'}(\varphi)<0$ as a result $V(\varphi)\propto\varphi^{-n}$.\\
The third case: if $H$ is negative then there is no condition for potential, therefore we suppose $V(\varphi)\propto\varphi^{-n}$. The pointed potential gives us opportunity to study the cosmology evolution of equation of state for the proposed model.\\
In order to explore the possibility of $\omega$ across -1, we have to check out $\frac{d}{dt}(p+\rho)\neq 0$ when $\omega\neq -1$,
\begin{equation}\label{32}
\frac{d}{dt}(p+\rho)=\frac{4}{3}(H\varphi+\dot{\varphi})(\dot{H}\varphi+H\dot{\varphi}+\ddot{\varphi})+\frac{\sqrt{2}}{3}\lambda_{2}^{2}(\dot{H}\varphi+H\dot{\varphi}+
\ddot{\varphi})+\frac{\lambda_{1}^{2}}{3}(V^{''}(\varphi)\dot{\varphi}\varphi+V^{'}(\varphi)\dot{\varphi}),
\end{equation}
by assuming $\dot{\varphi}=\ddot{\varphi}=0$, one can rewrite following equation,
\begin{equation}\label{33}
\frac{d}{dt}(p+\rho)=\frac{4}{3}(H\varphi)(\dot{H}\varphi)+\frac{\sqrt{2}}{3}\lambda_{2}^{2}(\dot{H}\varphi),
\end{equation}
therefore $\frac{d}{dt}(p+\rho)\neq0$, otherwise $H\varphi=-\frac{\sqrt{2}}{4}\lambda_{2}^{2}$ (it means that $H\varphi$ is constant). This is not true because $H\varphi$ is function of time. So, the above result proves the equation of state cross -1, therefore we have bouncing universe. The equation of state given by,
\begin{equation}\label{EoS}
\omega=\frac{\frac{(H\varphi)^{2}}{6}+\frac{\sqrt{2}}{3}\lambda_{2}^{2}H\varphi-\frac{\lambda_{1}^{2}}{\varphi^{2}}
-\lambda_{1}^{2}V(\varphi)}{\frac{(H\varphi)^{2}}{6}+\lambda_{1}^{2}V(\varphi)}.
\end{equation}
In the plots of the Fig. 1 we draw equation of state parameter in terms of time to see effect of deformation parameters separately. In the left plot we can see the effect of $\beta$ while in the right plot we can see the effect of $\beta$.\\

\begin{figure}[h]
\hspace*{1cm}
\begin{center}
\epsfig{file=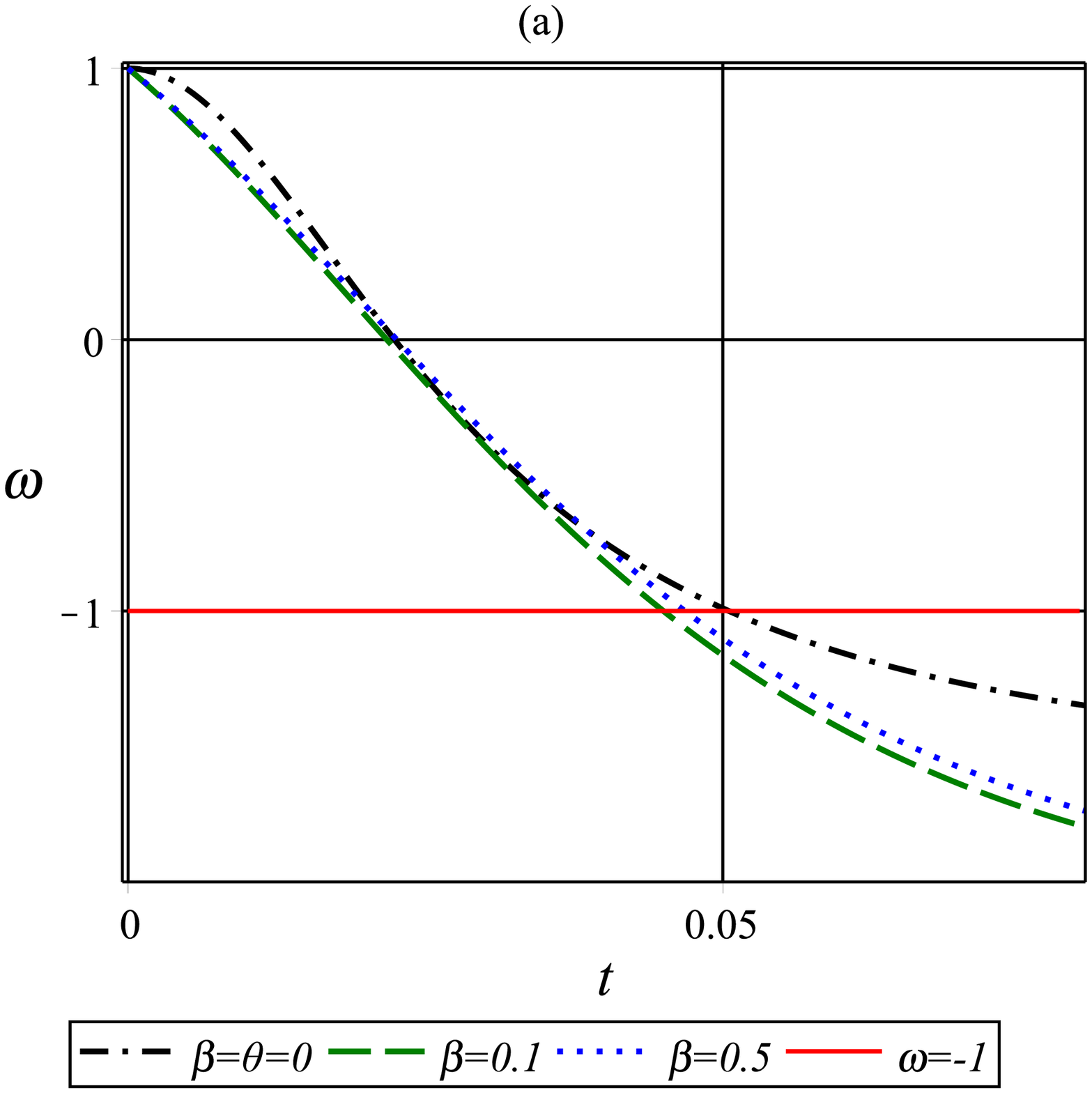,width=7cm}\epsfig{file=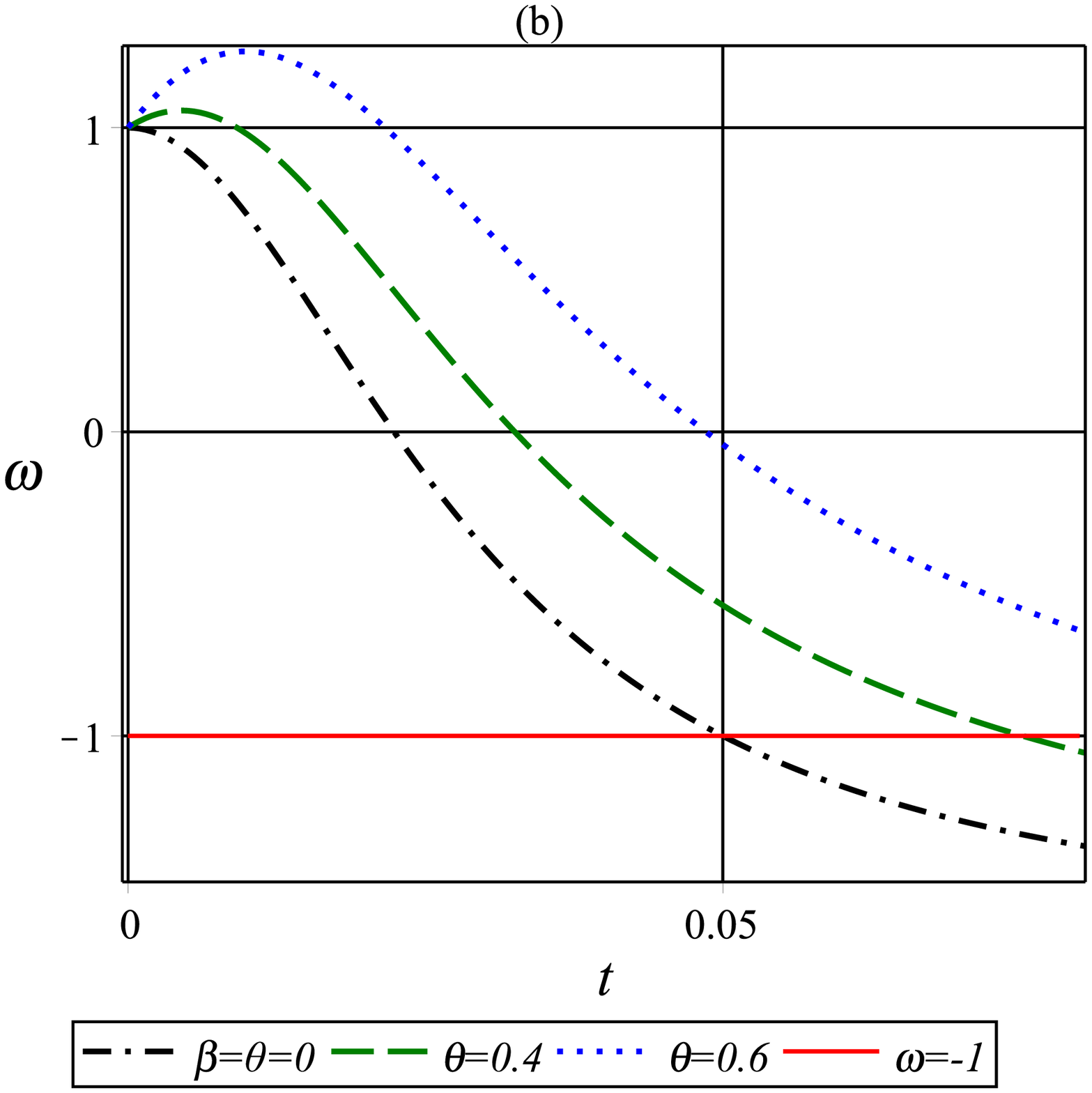,width=7cm}
 \caption{The variation $\omega$ with respect to $t$, where $V(\varphi)=\frac{V_{0}}{2\varphi^{2}}$, $\varphi=0.5$, $V_{0}=0.5$. (a)  $\theta=2$; (b) $\beta=0.1$}
\end{center}
\end{figure}

It is illustrated that the variation with $\beta$ is infinitesimal while increasing $\theta$ increases value of the $\omega$ (see Fig. 2 (b)).
We can see from Fig. 2 that the equation of state parameter yields to a constant at the late time. It is shown that in presence of the deformation parameters, the equation of state parameter across -1 at earlier time. The values of parameters fixed according to the stability condition which discussed below.\\
Here, also we can check the stability of system due to deformation of the non-minimally coupled to gravity model. In order to examine the stability of system we need to obtain the variation of pressure with respect to density energy of system. Hence, we investigate the stability of non-minimal coupling to gravity with deformation phase space variables. In that case we want to consider the stability by useful functions as $C_{s}^{2}=\frac{dp}{d\rho}$. In order to have stability with non-minimal deformation model such function must become more than zero. So, if we want to investigate the stability of system the above mentioned function can be expressed by sound speed in prefect liquid \cite{63}. By using the equations (\ref{26}), (\ref{29}) and $H^{2}=\frac{\rho}{3}$, one can obtain the sound speed,
\begin{eqnarray}\label{34}
\frac{dp}{d\rho}&=&\frac{\frac{1}{3}(H\varphi+\dot{\varphi})(\dot{H}\varphi+H\dot{\varphi}+\ddot{\varphi})+\frac{\sqrt{2}}{3}\lambda_{2}^{2}(\dot{H}\varphi+H\dot{\varphi}+\ddot{\varphi})}
{(H\varphi+\dot{\varphi})(\dot{H}\varphi+H\dot{\varphi}+\ddot{\varphi})+\lambda_{1}^{2}V^{'}(\varphi)\dot{\varphi}}
\nonumber\\
&+&\frac{\frac{\lambda_{1}^{2}}{3}(V^{''}(\varphi)\dot{\varphi}\varphi+V^{'}(\varphi)\dot{\varphi})-\lambda_{1}^{2}V^{'}(\varphi)\dot{\varphi}}
{(H\varphi+\dot{\varphi})(\dot{H}\varphi+H\dot{\varphi}+\ddot{\varphi})+\lambda_{1}^{2}V^{'}(\varphi)\dot{\varphi}}.
\end{eqnarray}
If we assume $\dot{\varphi}=\ddot{\varphi}=0$, then $c_{s}^{2}\equiv C_{s}$ lead us to have following condition,
\begin{equation}\label{35}
\frac{1}{3}+\frac{\sqrt{2}}{3}\frac{\lambda_{2}^{2}}{H\varphi}>0.
\end{equation}
So, the above condition for the stability of non-minimally coupled to gravity in case of deformed phase space lead us to have $4\theta > \beta$.  Here, we draw the variation of pressure with respect to the variation of energy density in terms of time ($C_{s}=\frac{dp}{d\rho}$) and show that how the stability correspond to two parameters $\theta$ and $\beta$ from NC geometry. It is illustrated by the Fig. 3 that non-deformed case is completely stable as well as the deformed case with condition $\beta=4\theta$ (see solid red line of the Fig . 3).

\begin{figure}[h]
\hspace*{1cm}
\begin{center}
\epsfig{file=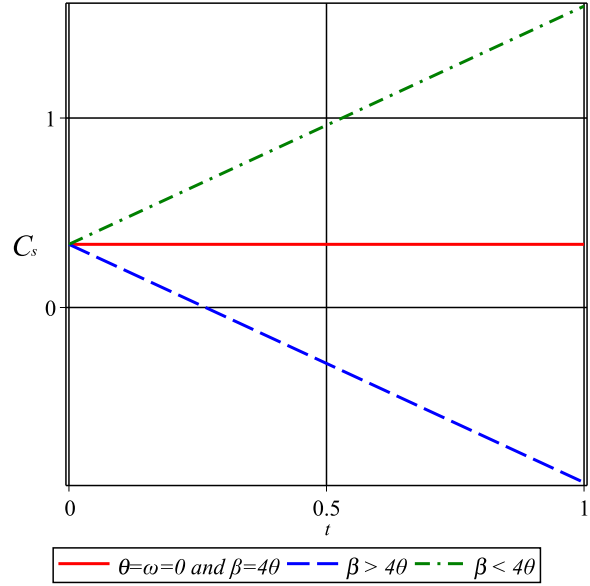,width=8cm}
 \caption{The variation $\frac{dp}{d\rho}$ with respect to $t$, where $V(\varphi)=\frac{V_{0}}{2\varphi^{2}}$, $\varphi=0.5$, $V_{0}=0.5$, $\theta=0.5$. }
\end{center}
\end{figure}

\section{Conclusion}\label{sec5}
In this paper, we introduced non-minimally coupled gravity and wrote Lagrangian. Also, by using some suitable variables, we obtained new Hamiltonian and Lagrangian. In order to deform the above mentioned system we arranged the harmonic oscillator form of Hamiltonian. We used the NC geometry and deformed
the new Hamiltonian and Lagrangian. Now, we take advantage from dark energy solution and investigated bouncing universe in deformed non-minimal coupling to gravity model. The condition of bouncing universe in deformed model lead us to arrange the potential in equation of state. In order to have bouncing solution we drew the equation of state in terms of time. It shown that the equation of state crossing $-1$. Also, we obtained some suitable relation between the $\theta$ and $\beta$  parameters of NC geometry to satisfy by bouncing universe. We found special condition ($\beta=4\theta$) where the stability of deformed model behaves like non-deformed case. In the case of $\beta\leq4\theta$ the deformed model is completely stable. Otherwise the model is initially stable which yields to unstable phase at the late time. Hence it is possible to have stable/unstable phase transition.
 In order to have solution of dark energy, we arranged the potential in the equation of state of deformed non-minimally coupled to gravity. In that case we draw the equation of state in terms of time and shown that the equation of state cross $-1$. Finally, we employed the pressure and energy density and investigated the stability of deformed non-minimally coupled to gravity. Here, we obtained the relation between $\theta$ and $\beta$ in NC for the approving the stability of system. Also here, we had some figure for the describing of stability with some special values of parameters of NC geometry. The relation between the above deformation of corresponding model and Finsler geometry will be important in future  for the describing some parameter in physics. Also, such relation give us more degree of freedom to compare the obtained results of theory with data. As we know the results from Finsler geometry have a relation with non-commutative geometry. So, we hope to  continue this research for the different cosmological model in future. For instance, recently bouncing universe considered in the contexts of generalized cosmic Chaplygin gas and variable modified Chaplygin gas, now it is interesting to consider bouncing universe in the context of extended Chaplygin gas \cite{ECG1, ECG2, ECG3, ECG4, ECG5} and use method of this paper. It is also possible to use the same method of this paper to study the holographic Barrow dark energy model \cite{end1, end2}.

\end{document}